\documentclass[12pt]{article}

\usepackage{graphics}
\usepackage{amssymb}
\usepackage{amsthm}
\usepackage{amsmath}
\usepackage{graphicx}
\usepackage[nodots,nocompress]{numcompress}

\begin{document}




\title{A Stochastic Finite Element Model for the Dynamics of Globular Macromolecules}


\author{$^*$ Robin Oliver, $^+$Daniel J Read, $^+$ Oliver G Harlen, $^*$ Sarah A Harris\\$^*$ School of Physics and Astronomy,\\ $^+$School of Mathematics,\\ University of Leeds, Leeds, LS2 9JT, UK}

\maketitle
\begin{abstract}
We describe a novel coarse-grained simulation method for modelling the dynamics of globular macromolecules, such as proteins. The macromolecule is treated as a continuum that is subject to thermal fluctuations. The model includes a non-linear treatment of elasticity and viscosity with thermal noise that is solved using finite element analysis. We have validated the method by demonstrating that the model provides average kinetic and potential energies that are in agreement with the classical equipartition theorem. In addition, we have performed Fourier analysis on the simulation trajectories obtained for a series of linear beams to confirm that the correct average energies are present in the first two Fourier bending modes. We have then used the new modelling method to simulate the thermal fluctuations of a representative protein over 500ns timescales.
Using reasonable parameters for the material properties, we have demonstrated that the overall deformation of the biomolecule is consistent with the results obtained for proteins in general from atomistic molecular dynamics simulations.
\end{abstract}






\section{Introduction}

The conformational dynamics of biological macromolecules poses a unique challenge to computational physicists. Proteins are chemically inhomogeneous and aperiodic. Even small proteins can contain many thousands of atoms whereas molecular motors, such as the ribosome, contain many hundreds of thousands of atoms [1].  Many active biological structures, such as the transcriptional machinery, are large protein complexes made up from numerous (of order 10) separate proteins loosely bound by non-covalent interactions [2]. Biomolecules are soft nanoscale objects. They exhibit large conformational changes both due to thermal fluctuations and interactions with other biomolecules, which are often critical to their function [3]. Most remarkably, motor proteins are capable of exerting mechanical force to produce motility both at the level of individual molecules, or, when acting co-operatively, at the macroscopic level [4]. 
Biomolecular dynamics spans an enormous range of timescales, from the vibration of individual atoms or groups of atoms over picosecond timescales to the action of a molecular motor, which may require milliseconds.

As a result of this complexity, computer simulation studies of the conformational dynamics of biomolecules and the interactions within biomolecular complexes are invaluable for interpreting experimental data and for probing the physical mechanisms used by proteins to perform their function. However, even with the advent of parallel supercomputing, the most established techniques for simulating biomolecular dynamics are limited by their computational expense. Calculations at the fully quantum mechanical level, which are capable of resolving electronic structure, are generally limited to static fragments of individual proteins [5]. Simulations in which proteins and their solvent environment are described in atomistic detail have reached timescales in excess of 1 microsecond [6], which is sufficient to capture vital biological processes such as the folding of small proteins. 
For larger biomacromolecules, such as the ribosome, accessible simulation timescales are reduced to nanoseconds [7], which is considerably shorter than the millisecond timescales over which the molecular motor operates. Moreover, molecular biology occurs in a cellular environment that is densely packed with proteins and membrane surfaces enclosing higher order cellular structures. If computer models are to be used in a truly \emph{in vivo} context then, {we will need to move beyond computer models that consider only fragments of proteins or single proteins.}


Strategies for reducing the computational cost of protein dynamics simulations include simplifying the force-field to exclude anharmonic terms (so that the conformational fluctuations of the protein can be calculated by normal mode analysis), reducing the number of particle interactions that need to be considered by coarse-graining a subset of atoms into one entity, or a combination of these approaches. A comprehensive review of coarse-grained protein modelling has recently been published by Tozzini [8]. {Using these lower resolution coarse-grained techniques, simulation time and length scales have been extended when compared with fully atomistic molecular dynamics}. Most notably, McGuffie and Elcock were recently able to simulate the bacterial cytoplasmic environment using Brownian dynamics simulations and a rigid protein model [9]. The simulation contained over 1000 proteins, and captured 20 $\mu$s of dynamics.


The majority of computer simulation studies of conformational change in proteins consider the macromolecule to be constructed from discrete particles, where a particle may represent one (as in atomistic simulation) or more atoms. Although a `particle' description is appropriate for atomically detailed calculations, at the coarse-grained level the particle size and the number of atoms that they represent become arbitrary, and are typically chosen for computational convenience. At the mesoscale (length scales from hundreds of nanometers to microns), hybrid fluid mechanics/solid mechanics techniques, such as the Immersed Boundary Method (IBM) [10] [11] and the Immersed Finite Element Method (IFEM) [12, 13] have been developed to model how the shapes or positions of objects change in response to hydrodynamic fluctuations and fluid flow [14]. 
These techniques have been used, for example, to study the stochastic desorption of rigid nanoparticles from membrane surfaces under shear flows [15], the Brownian motion of nanoparticles in a Newtonian fluid due to thermally induced hydrodynamic fluctuations [16, 17] and the deformation of vesicle being dragged by a Brownian ratchet model of a biological motor protein [18]. The IBM treats an object immersed in a fluid as a series of particles that interact with both with one another and the background fluid, which is placed on a grid [10]. In IFEM, both the object and the fluid are represented on two separate meshes which are superimposed. Thermal fluctuations are introduced by subjecting the mesh to a fluctuating stress using the appropriate Langevin equations [13].

Here we describe a continuum model for globular macromolecules which is designed to model the conformational dynamics of individual large proteins and biomolecular complexes. Since it is a continuum model, it cannot provide atomic resolution information, which places a lower bound on the molecular sizes that it can consider. However, it has no upper size limit, meaning that it is capable of taking biomolecular simulation from the atomistic into the mesoscale.  The {viscous and elastic} nature of biological matter at the mesoscale was recently demonstrated experimentally [19]. Therefore, rather than using discrete inter-particle potentials to represent interactions within the biomolecule, our new algorithm uses material quantities, namely the density, bulk/shear moduli and bulk/shear viscosities to describe the mechanics of the protein, and evolves the shape of the biomolecule in response to stress using FEA. 
We have developed a numerical scheme that includes a locally calculated fluctuating stress to account for thermal noise; we refer to this new scheme as Fluctuating Finite Element Analysis (FFEA). The concept of a fluctuating stress is not in itself new. It was first proposed by Landau [20] and has subsequently been utilised by Sharma and Patankar [21] to solve for the Brownian motion of particles by including a fluctuating stress in the fluid. In FFEA, we incorporate the fluctuating stress tensor directly into the biological material which then deforms due to thermal agitation. The technique has the additional advantage over particle-based simulation methods that it does not require an atomically detailed experimental structure as input to the calculations. Rather, the continuum model can use lower resolution structural data, so long as the overall shape of the protein is known.

We present the mathematical background to FFEA in Section 2, and demonstrate its consistency with the Fluctuation Dissipation Theorem. In Section 3, we validate FFEA by simulating the deformation of a simple rod and comparing with the expected analytical result. In Section 4, we demonstrate the application of FFEA to model a protein for which only low resolution structural data exists.

\section{Mathematical Background}

\subsection{The Continuum Model}Fluctuating Finite Element Analysis (FFEA) treats a macromolecule as a continuous medium of density $\rho$ subject to thermal noise, viscous dissipation and elasticity. The material is described by a Kelvin-Voigt model, where the viscous stresses and elastic stresses are added together. Hydrodynamics external to the macromolecule are neglected in the current model, however internal hydrodynamics are included. The equation of motion connecting the velocity $u_i$ to the stress $\sigma_{ij}$ at all points in the material can be represented by continuum fields. Using indices to refer to spatial direction, the equation of motion is then:

\begin{eqnarray}
\rho \left ( \frac{\partial u_i}{\partial t} +u_{j}\frac{\partial u_i}{\partial x_j} \right ) &=& \frac{\partial \sigma_{ij}}{\partial x_j}
\end{eqnarray}

\noindent where $\left ( \frac{\partial u_i}{\partial t} +u_{j}\frac{\partial u_i}{\partial x_j} \right )$ is the total time derivative of the velocity vector field in the Lagrangian frame of the material. The stress $\sigma_{ij}$ can be subdivided into three contributions:

\begin{equation}
\sigma_{ij}=\sigma^v_{ij}+\sigma^e_{ij}+\sigma^t_{ij}
\end{equation}

\noindent $\sigma^v_{ij}$, $\sigma^e_{ij}$ and $\sigma^t_{ij}$ are the stresses due to non-conservative friction, elastic conservative forces and thermal fluctuations respectively. Although we use a {Kelvin-Voigt material model} for deriving these stresses in the current work, our method is not limited to this treatment; more sophisticated (and more realistic) material models could be incorporated without altering the basic structure of the method. {However, for material models with fading memory ensuring that the thermal noise derived in Section 2.3.1 and 2.3.2 correctly obeys the fluctuation-dissipation theorem is more complex}. We now describe the form of the three elastic stress terms  we have used (Section 2.1.1-2.1.3) and the nature of the finite element approximation employed (in Section 2.2). Finally, in Section 2.3 we provide additional details of the thermal noise term $\sigma^t_{ij}$ 
and demonstrate the compliance of the fluctuating finite element scheme with the fluctuation-dissipation theorem.

\subsubsection{Viscous Stress} The material is assumed to have an isotropic linear viscous stress $\sigma^v_{ij}$ which can be written as:

\begin{eqnarray}
\sigma^v_{ij} &=& \mu \left ( \frac{\partial {u}_i}{\partial {x}_j} +\frac{\partial {u}_j}{\partial {x}_i} \right ) + \lambda \frac{\partial {u}_m}{\partial {x}_m}\delta_{ij},
\end{eqnarray}

\noindent where $\mu$ is the shear viscosity and $\lambda$ is the second coefficient of viscosity, giving a bulk viscosity $\mu_{bulk}=\lambda+\frac{2}{3}\mu$.

\subsubsection{Elastic Stress} In this paper we consider the simple case where the material described is hyperelastic, so that the elastic stress $\sigma^e_{ij}$ can be derived from a strain energy density functional. This is written in terms of the deformation gradient tensor $\underline{F}$ (defined as $F_{ij}=\frac{\partial x_i}{\partial X_j}$ where $\underline{x}(\underline{X},t)$) is the current position of material initially located at $\underline{X}$.  Hence the local volume change is given by $\frac{V}{V_0}=det(\underline{F})$.  We use a formulation which includes classical rubber elasticity and a volumetric spring that acts as a source of pressure. The strain energy density per unit current volume is written as:

\begin{eqnarray}
W&=&\frac{G}{2 det(F)} tr(F F^T) +\frac{B}{2 det(F)} \left (det(F) -\alpha \right )^2 \nonumber \\ &-&\frac{3G}{2det(F)}-\frac{B}{2det(F)}\left (\frac{G}{B} \right )^2.
\end{eqnarray}

\noindent Here G is the shear modulus, $B-\frac{G}{3}$ is the bulk modulus K, $\underline{F}$ is the deformation gradient tensor and $\alpha$ is a constant used to impose zero isotropic stress at zero deformation, requiring that $ \alpha=1+\frac{G}{B}$. {From the bulk and shear moduli, the Young's Modulus of the material is given by  $E=\frac{9KG}{3K+G}$.}

The stress can be calculated by considering the the change in energy when a small strain $\epsilon_{ij}$ is applied to a portion of material, giving the stress tensor:

\begin{eqnarray}
\sigma^e_{ij}&=&\frac{1}{det(F)}\frac{\partial (W det(F))}{\partial \epsilon_{ij}} \nonumber \\
&=&\frac{G}{det(F)}F_{ik}{F^T}_{kj}+ B(det(F)-\alpha)\delta_{ij}.
\end{eqnarray}

{We have assumed that the effect of the thermal noise on the elasticity of the material is small compared to the uncertainty in the known elastic moduli for biomaterials (see Section 4). In general, small length scale thermal fluctuations will indeed affect the effective elasticity over larger length scales in the non-linear elastic regime. In principle, this effect should be accounted for when coarse-graining if the dimensions of the finite elements are significantly increased.}

\subsubsection{Thermal Stress} In particle based simulation techniques such as Molecular Dynamics or Brownian Dynamics (BD) thermal fluctuations are included by adding a random force to each particle in the simulation. In our method, thermal forces are introduced via a fluctuating stress tensor $\sigma^t_{ij}$. Unlike the elastic and viscous contributions, this thermal stress term is stochastic in both space and time, with statistics chosen to balance the viscous energy dissipation. The advantage of this approach is that in a finite element approximation the fluctuating stress can be calculated entirely locally for the viscous stress and still yield the correct thermal physics. In Section 2.3 we derive the fluctuation dissipation relation for this model and show that at equilibrium the input of energy into the system by the noise and the reduction of energy from the viscous terms do indeed balance appropriately, and consequently that the fluctuation dissipation theorem is
satisfied.

\subsection{Finite Element Approximation} In order to construct a finite element discretisation, we seek a weak form of Equation (1) by performing a volume integral with the weight function $w$(\textbf{x}) to give:

\begin{eqnarray}
\int_V \rho w(\textbf{x}) \frac{\partial u_i}{\partial t} dV = -\int_V \frac{\partial w(\textbf{x})}{\partial x_j}\sigma_{ij} dV+ \int_S f_i w(\textbf{x}) dS.
\end{eqnarray}

\noindent where $f_i$ are the external surface fraction forces. This corresponds to the standard application of the finite element method [22]. The {second} integral in Equation (6) can now be evaluated by substituting in the components of the stress.

\begin{eqnarray}
\int_V \frac{\partial w(\textbf{x})}{\partial x_j}\sigma_{ij} dV&=& \int_V \left ( \mu \frac{\partial w(\textbf{x})}{\partial {x}_{j}} \frac{\partial u_{i}}{\partial {x}_{j}} + \mu  \frac{\partial w(\textbf{x})}{\partial {x}_j}\frac{\partial u_j}{\partial {x}_i} + \lambda \frac{\partial w(\textbf{x})}{\partial {x}_i} \frac{ \partial u_{j}}{\partial {x}_j} \right ) dV \nonumber \\  && {} + \int_V \frac{\partial w(\textbf{x})}{\partial x_j}\sigma^e_{ij} dV \nonumber \\ && {} + \int_V \frac{\partial w(\textbf{x})}{\partial {x}_j} \sigma^t_{ij}  dV.
\end{eqnarray}

Equation (7) contains first order derivatives of both the velocity vector $u_i$ and the the weight function $w(\textbf{x})$. Thus, both the functions ${u}_i$ and $w(\textbf{x})$ must be differentiable over the domain of the differential equation and square integrable. Therefore, a suitable space is such that ${u}_i, w(\textbf{x})\in  H^1_0(\omega)$ where $\omega$ is the domain of the differential equation in 3-space, while the tensors $\sigma^e_{ij}$ and $\sigma^t_{ij}$  can be defined as $\sigma^e_{ij}, \sigma^t_{ij}\in  L_2$.

With the solution space now defined, we subdivide the domain $\omega$ of the differential equation into finite elements with nodes that are fixed in the Lagrangian frame of the material so that the velocity is expressed in the form $u_i=\sum_{\alpha} v_{i \alpha} \phi_\alpha$ where $\phi_{\alpha}$ are base vectors that span the subspace of $H^1_0(\omega)$ so that,

\begin{equation}
\frac{Du_i}{D t} = \sum_{\alpha} \frac {\partial v_{i \alpha}}{\partial t} \phi_{\alpha}.
\end{equation}

\noindent So, Equation(6) becomes:

\begin{eqnarray}
{M}_{pq}\frac{\partial {v}_{q}}{\partial t}+{K}_{pq} {v}_{q}&=&{E}_{p} +{N}_{p},
\end{eqnarray}

\noindent where $M_{p(i,\beta) q(j,\alpha)}$, $K_{p(i,\beta) q(j,\alpha)}$, $E_{p(i,\beta)}$ and $N_{p(i,\beta)}$ are defined below. This treatment corresponds to the Galerkin formulation [22] of finite element analysis where the weight functions $w(\textbf{x})$ are chosen to be the same as the basis functions $\phi_{\alpha}$. We have also introduced the indices $p$ and $q$ as counting indexes for the full dimension of the finite element system such that $p$ can be written as $p(i,\beta)$ and $q$ can be written as $q(j,\alpha)$. This gives:

\begin{eqnarray}
{M}_{p(i,\beta) q(j,\alpha)}&=&  \delta_{ij}\left (\int_V \rho \phi_{\alpha} \phi_{\beta} dV \right ), \\ {K}_{p(i,\beta) q(j,\alpha)}&=& \int_V \mu \frac{\partial \phi_{\beta}}{\partial {x}_c} \frac{\partial \phi_{\alpha}}{\partial {x}_{c}} \delta_{ij} + \mu \frac{\partial \phi_{\beta}}{\partial {x}_j}\frac{\partial  \phi_{\alpha}}{\partial {x}_i} + \lambda \frac{\partial \phi_{\beta}}{\partial {x}_i} \frac{\partial \phi_{\alpha}}{\partial {x}_j} dV, \\ {E}_{p(i,\beta) q(j,\alpha)}&=&-\int_V \frac{\partial \phi_{\beta}}{\partial {x}_j}\sigma^e_{ij} dV, \\ {N}_{p(i,\beta) q(j,\alpha)}&=& -\int_V \frac{\partial \phi_{\beta}}{\partial {x}_j} \sigma^t_{ij} dV.
\end{eqnarray}

Equation (9) describes a linear system of Langevin equations that can be solved for $\frac{\partial {v}_{q}}{\partial t}$ by inverting the the mass matrix $M_{pq}$. Physically, the different matrices presented in Equation (9) describe each of the particular processes that govern the behaviour of a macromolecule in FFEA. The mass matrix $M_{pq}$ describes how mass is distributed throughout the finite elements, $K_{pq}$ describes how the model dissipates energy through viscosities,  $E_{p}$ is an elastic force vector and $N_{p}$ is a thermal force vector.

\subsection{Thermal Noise}

The remaining undefined quantity in Equation (9) is the fluctuating stress tensor $\sigma^t_{ij}$. To derive the form of $\sigma^t_{ij}$ we first derive the fluctuation dissipation relation for this system.

{For the case of a Kelvin-Voigt material, the derivation of the fluctuation-dissipation theorem is simplified because the elastic stress in the model is derived from a strain energy that depends only upon the instantaneous deformation of the system. Consequently, the elastic terms in this model are conservative and the energy stored during a structural distortion is not dissipated by the material. By contrast, in material models with fading memory (such as the Maxwell model), the viscoelastic stress is dependent on the strain history of the material. In such cases, there will be additional dissipation of energy due to memory effects within the material, and the derivation of the corresponding fluctuation-dissipation theorem is less straightforward.}

\subsubsection{Fluctuation Dissipation Theorem} The overall equation of motion of the macromolecule is comprised of a linear system of Langevin equations. Deriving the fluctuation dissipation relation for this specific system is necessary to provide the statistics of $N_{p}$.  We can re-write Equation (9) as:

\begin{equation}
{M}_{p \alpha} \frac{ \partial {v}_\alpha}{\partial t} = {N}_p - K_{p \gamma} {v}_{\gamma} - \nabla_p U(\textbf{x}).
\end{equation}

\noindent where the elastic vector $E_p$ has been re-written in the form $E_p=-\nabla_p U(\textbf{x})$, such that the potential $U$ is the strain energy. We have relabeled the dummy indices from $q$ to $\alpha$ and $\gamma$ for clarity. We use the summation convention in Equation (14) throughout this section.

{The derivation of the fluctuation dissipation theorem first considers the total kinetic energy $E_k$ of the system:}

\begin{equation}
{{E}_{k}=\frac{v_{\alpha}M_{\alpha \beta} v_{\beta}}{2}.}
\end{equation}
{Equation (15) is exact within the discretised finite element framework. The probability of finding the system with a given given kinetic energy is therefore proportional to:}
\begin{equation}
{P \sim \exp(-\frac{v_{\alpha}M_{\alpha \beta} v_{\beta}}{2k_B T}).}
\end{equation}
{Since this is a generalised normal distribution, it follows that the second moment average of the node velocities must, at equilibrium, be:}
\begin{equation}
{{Q}_{pq}=\langle {v}_p {v}_q \rangle = k_b T {M}^{-1}_{pq},}
\end{equation}
{which is the equipartition theorem for this system. Equation (17) is exact within the discretised finite element framework, so the fluctuation dissipation theorem derived from it is also exact. However, in practice numerical errors will occur due to the numerical integrator, as is discussed in Section 3.1.1.} The derivation of the fluctuation dissipation relation for this system follows by analyzing fluctuations in the energy variable ${Q}_{pq}$, considering its change $\Delta Q_{pq}$ during time step $\Delta t$ (with the intention of letting $\Delta t$ become infinitesimally small) such that:

\begin{equation}
\Delta {Q}_{pq}= \langle \Delta {v}_p {v}_q +{v}_p \Delta {v}_q +\Delta {v}_p \Delta {v}_q \rangle=0,
\end{equation}

\noindent where from Equation (14):

\begin{equation}
\Delta {v}_p= \Delta t {M}^{-1}_{p \alpha}({N}_{\alpha} -{K}_{\alpha \gamma} {v}_{\gamma} + \nabla_{\alpha} U(\textbf{x})).
\end{equation}

\noindent Thus, the terms on the right hand side of Equation (18) are given by:

\begin{equation}
\Delta {v}_p {v}_q = \Delta t {M}^{-1}_{p \alpha}({N}_{\alpha} -{K}_{\alpha \gamma} {v}_{\gamma} + \nabla_{\alpha} U(\textbf{x})){v}_q,
\end{equation}

\begin{equation}
{v}_p \Delta {v}_q = {v}_p \Delta t {M}^{-1}_{q \delta}({N}_{\delta} -{K}_{\delta \epsilon} {v}_{\epsilon} + \nabla_{\delta} U(\textbf{x})),
\end{equation}
and finally:
\begin{equation}
\Delta {v}_p \Delta {v}_q = \Delta t^2 {M}^{-1}_{p \alpha}({N}_{\alpha} -{K}_{\alpha \gamma} {v}_{\gamma} + \nabla_{\alpha} U(\textbf{x})){M}^{-1}_{q \delta}({N}_{\delta} -{K}_{\delta \epsilon} {v}_{\epsilon} + \nabla_{\delta} U(\textbf{x})).
\end{equation}

To evaluate the ensemble averages of Equations (20), (21) and (22) to order $\Delta t$ we note $\langle v_p \rangle =0$, and that:

\begin{equation}
\langle M^{-1}_{p \alpha} \nabla_{\alpha} Uv_q +M^{-1}_{q\delta} \nabla_\delta Uv_p \rangle=0.
\end{equation}

 \noindent Because at equilibrium the total energy of the system is simply the sum of the kinetic and potential energies, the probability of finding the system in any given microstate is also the product of the probability distributions describing the range of potential and kinetic energies the system can adopt. Therefore, the potential and velocity terms are uncorrelated at equilibrium. Since the kinetic energy contains only terms quadratic in $v_p$, it follows that $\langle v_p \rangle=0$, so Equation (23) must hold. Equations (20)-(22) then simplify to:

\begin{eqnarray}
\langle \Delta {v}_p {v}_q \rangle &=& -\Delta t {M}^{-1}_{p \alpha} {K}_{\alpha \gamma} \langle {v}_{\gamma} {v}_q \rangle \nonumber \\ &=& -\Delta t k_B T {M}^{-1}_{p \alpha} {M}^{-1}_{\gamma q} {K}_{\alpha \gamma},
\end{eqnarray}

\begin{eqnarray}
\langle  {v}_p \Delta {v}_q \rangle &=& -\Delta t {M}^{-1}_{p \delta} {K}_{\delta \epsilon} \langle {v}_{p} {v}_{\epsilon} \rangle \nonumber \\ &=& -\Delta t k_B T {M}^{-1}_{q \delta} {M}^{-1}_{p \epsilon} {K}_{\delta \epsilon},
\end{eqnarray}

\begin{eqnarray}
\langle \Delta {v}_p \Delta {v}_q \rangle = \Delta t^2 {M}^{-1}_{p \alpha} {M}^{-1}_{q \delta} \langle {N}_{\alpha} {N}_{\delta} \rangle.
\end{eqnarray}
Direct substitution of Equations (24)-(26) into (18) leads to the following,

\begin{eqnarray}
\Delta t^2 \langle {M}^{-1}_{p \alpha} {N}_{\alpha} {M}^{-1}_{q \delta} {N}_{\delta} \rangle &=&\Delta t k_B T ( {M}^{-1}_{p \alpha} {K}_{\alpha \gamma} {M}^{-1}_{\gamma q} +{M}^{-1}_{q \delta} {K}_{\delta \epsilon} {M}^{-1}_{p \epsilon}).
\end{eqnarray}

\noindent Multiplying through by the mass matrix gives:

\begin{eqnarray}
\Delta t^2 \langle \delta_{p \alpha} {N}_{\alpha} \delta_{q \delta} {N}_\delta \rangle &=& \Delta t k_B T ( \delta_{p \alpha} \delta_{\gamma q} {K}_{\alpha \gamma} + \delta_{q \delta} \delta_{p \epsilon} {K}_{\delta \epsilon}),
\end{eqnarray}

\noindent and so,

\begin{eqnarray}
\langle {N}_p {N}_q \rangle &=& \frac{k_B T}{\Delta t}({K}_{pq}+{K}_{qp})
\end{eqnarray}

\noindent which is the fluctuation dissipation relation for the equation of motion in Equation (14).

\subsubsection{Fluctuation Dissipation Relation for Linear Elements} In order to solve Equation (29) and derive the nature of the thermal stress tensor $\sigma^t_{ij}$, we must choose a set of basis functions $\phi_{\alpha}$. The simplest choice of basis functions are those of a linear tetrahedron. Equation (11) provides an explicit expression for the viscous matrix ${K}_{pq}$:

\begin{eqnarray}
{K}_{pq}&=& \int_V \mu \frac{\partial \phi_{\beta}}{\partial {x}_c} \frac{\partial \phi_{\alpha}}{\partial {x}_{c}} \delta_{ij} + \mu \frac{\partial \phi_{\beta}}{\partial {x}_j}\frac{\partial  \phi_{\alpha}}{\partial {x}_i} + \lambda \frac{\partial \phi_{\beta}}{\partial {x}_i} \frac{\partial \phi_{\alpha}}{\partial {x}_j} dV.
\end{eqnarray}

 In the case of linear elements the derivatives of the basis functions are constants and Equation (29) can easily be simplified. In order to satisfy the fluctuation dissipation relation, we must assign an appropriate form to the fluctuating stress tensor $\sigma^t_{ij}$ {that is $\delta$-correlated in space and time}. Firstly, $\sigma^t_{ij}$ must be symmetric such that $\sigma^t_{ij}=\sigma^t_{ji}$ and must consist of at least 7 independent stochastic processes. The solution we have found has a total of 7 distinct stochastic processes and is of the following form:

\begin{equation}
\sigma^t_{ij}= \left ( \frac{2k_BT}{V\Delta t}\right )^{\frac{1}{2}}\left ( \mu^{\frac{1}{2}} X_{ij}+\lambda^{\frac{1}{2}} X^0\delta_{ij}\right ).
\end{equation}

\noindent where $X_{ij}$ is a stochastic tensor containing 6 independent stochastic processes such that $X_{ij}=X_{ji}$ and $X^0$ is a stochastic variable independent of any variable in $X_{ij}$ such that:

\begin{eqnarray}
\langle X_{ij} \rangle &=& 0, \\ \langle X^0 \rangle &=& 0, \\ \langle X_{ij} X_{kl} \rangle &=& \delta_{ik}\delta_{jl}+\delta_{il}\delta_{jk}, \\ \langle X^0 X^0 \rangle &=& 1, \\ \langle X^p X_{ij}, \rangle &=&0.
\end{eqnarray}

\noindent Note that the correlation function for the shear noise in Equation (34) is equivalent to that used by Sharma and Patankar [21].

{The thermal stress tensor is $\delta$-correlated in both space and time.The spatial $\delta$-correlation is ensured by the finite element discretisation of the system, which guarantees that each element is independent of all the others. Since the viscous dissipation within a single element depends only upon the instantaneous deformation rate, there is no dependence on history of deformation and the fluctuations must also be $\delta$-correlated in time.}

We now show that this choice satisfies Equation (29) {and thus verify that for this Kelvin-Voigt material model the fluctuation dissipation theorem is obeyed}. Note that we also convert from the $p$ and $q$ notation back to $i, \beta$ and $j, \alpha$ so that the resulting matrices from Equation (29) can be directly compared.

\begin{eqnarray}
\langle {N}_{p(i,\beta)} {N}_{q(j,\alpha)} \rangle &=& \langle \int_V \frac{\partial \phi_{\beta}}{\partial {x}_c} \sigma^t_{ic}dV \int_V \frac{\partial \phi_{\alpha}}{\partial {x}_d} \sigma^t_{jd}dV \rangle \\ &=& V^2 \frac{\partial \phi_{\beta}}{\partial x_c} \frac{\partial \phi_{\alpha}}{\partial x_d}\langle \sigma^t_{ic} \sigma^t_{jd} \rangle
\end{eqnarray}

We now substitute Equation (31) into Equation (38), also using Equations (32)-(36), to yield the following:

\begin{eqnarray}
\langle {N}_{p} {N}_{q} \rangle &=& \left ( \frac{2 k_B TV}{\Delta t} \right ) \left ( \mu \frac{\partial \phi_{\beta}}{\partial x_c} \frac{\partial \phi_{\alpha}}{\partial x_c}\delta_{ij} +\mu \frac{\partial \phi_{\beta}}{\partial x_j} \frac{\partial \phi_{\alpha}}{\partial x_i}+ \lambda \frac{\partial \phi_{\beta}}{\partial x_i} \frac{\partial \phi_{\alpha}}{\partial x_j} \right ) \nonumber \\ &=& \left ( \frac{ k_B T}{\Delta t} \right ) \int_V 2\mu \frac{\partial \phi_{\beta}}{\partial x_c} \frac{\partial \phi_{\alpha}}{\partial x_c}\delta_{ij} +2\mu \frac{\partial \phi_{\beta}}{\partial x_j} \frac{\partial \phi_{\alpha}}{\partial x_i}+ 2\lambda \frac{\partial \phi_{\beta}}{\partial x_i} \frac{\partial \phi_{\alpha}}{\partial x_j} dV \nonumber \\ &=& \left ( \frac{ k_B T}{\Delta t} \right ) \left ( K_{pq} + K_{qp} \right )
\end{eqnarray}

\noindent where the factor of two arises because the viscosity matrix $K_{pq}$ is symmetric. This simple solution for the thermal stress tensor is valid only for linear elements because is assumes that the compression across an element is uniform.

For second (and higher) order elements, we have found it significantly less straightforward to obtain fluctuating stress terms satisfying the fluctuation dissipation relation. The difficulty arises because the derivatives of the basis functions are no longer constant, so the simple rearrangements in Equation (37) to (39), where the integrals are trivial, are no longer possible. In general, the fluctuating stress terms for second (and higher) order elements depend in a non-trivial manner on the shape of the element, making them impractical for our computational scheme.  In Section 3.1.5 below we present a simple scheme which allows elastic contributions to the stress to be treated using second-order elements, whilst retaining a first-order scheme for viscous and thermal stresses.

\section{The Numerical Method and its Validation} As discussed in Section 2 the finite element treatment of the stresses shown in Equation (2) results in an equation of motion for the system that is a linear system of Langevin equations:

\begin{equation}
{M}_{pq}\frac{\partial {v}_{q}}{\partial t}={K}_{pq}{v}_{q}+{E}_{p}+{N}_{p}.
\end{equation}

Equation (40) can then be numerically integrated using a standard time integrator such as Runge-Kutta (RK), velocity Verlet or Euler. An example of a simple loop to iteratively perform a time step is given below.

\begin{enumerate}
\item Characterise the initial conditions such as starting position and structure of the finite element mesh.
\item Calculate ${M}_{pq}$, ${K}_{pq}$, ${E}_{p}$, ${N}_{p}$ and $M^{-1}_{pq}$.
\item Evaluate the new velocity vector and node positions using a time integrator.
\end{enumerate}

The computational bottleneck in this loop is the calculation of the required matrices and vectors. However, that part of the algorithm can easily  be parallelised. The matrices are symmetric which reduces the computational load. Since the the mass matrix and its inverse remain constant through the simulation, they only need to be calculated once at the start of the simulation.

\subsection{Validation of the Continuum Model}
To validate the results derived in Section 2 and to demonstrate that the numerical algorithm reproduces the thermal physics of the system correctly, we have tested that the average kinetic and average potential energies converge to values required by the classical equipartition theorem for sufficiently small integration timesteps. To show that the method gives the expected changes in conformation of a thermally fluctuating mesoscopic object, we have used Euler Beam Theory to show that the average amount of potential energy found in the first two Fourier modes of a long beam also agrees with the equipartition value.

\subsubsection{Testing the Average Potential and Kinetic Energies} The average potential and kinetic energy depends on the number of degrees of freedom the system possesses. As this model only includes internal forces there is no solid body rotation or translation, effectively freezing out six degrees of freedom. Thus if $n$ is the number of nodes in the system, the total number of kinetic and potential degrees of freedom is $3n-6$. Therefore, from equipartition the average kinetic energy is given by:

\begin{eqnarray}
\langle E_{kin} \rangle  &=& \frac{(3n-6)k_BT}{2}.
\end{eqnarray}

If deformations are small then only harmonic terms in the elastic energy are important and the average potential energy becomes:

\begin{eqnarray}
\langle E_{pot} \rangle &=& \frac{(3n-6)k_BT}{2}.
\end{eqnarray}

The kinetic energy and potential energy for the system are then defined as follows:

\begin{eqnarray}
 E_{kin} &=& \frac{{v}_p {M}_{pq} {v}_q}{2}, \\ U&=& \sum_{\gamma}\int_{V_0}\frac{G}{2} tr(FF^T)^\gamma +\frac{B}{2} \left (det(F^{\gamma}) -\alpha \right )^2 dV_0 \nonumber \\ &&+ \sum_{\gamma}\int_{V_0}-\frac{3G}{2}-\frac{B}{2}\left (\frac{G}{B} \right )^2 dV_0.
\end{eqnarray}

\noindent Here the sum over $\gamma$ represents a sum over all the elements in the system and $V_0$ is the rest volume of each individual element.

Dimensionless systems were considered, in which the density, {viscosities, $\mu$ and $\lambda$ and elastic moduli $G$ and $B$ were set to unity throughout,} with $k_B T=0.0001$ to ensure small deformations. To test that the kinetic and potential energies comply with equipartition, we performed a simulation of a 54 element cylindrically meshed beam, which was constructed using the GMSH package [23]. The finite element mesh was firstly equilibrated from the initial configuration, and then the average kinetic and potential energies were calculated. The simulation was performed using 4 different time integrators; Euler, velocity Verlet, second order Runge-Kutta (R2) and forth order Runge Kutta (R4). Figure 1 shows the percentage error in the energies as a function of the integration time step. {The simulations were continued until the sampling errors in the average kinetic and potential energies were sufficiently small that the trends in Figure 1 could be clearly observed (this required 4 million timesteps for equilibration, and 20 million timesteps production run).} The simulation performed with all four integrators gives the correct equipartition value for short integration timesteps, indicating that the inclusion of thermal fluctuations into FEA provides the expected equipartition values for the kinetic and potential energy associated with the thermal fluctuations of the mesoscale beam.

When longer time steps are considered, R2 and R4 reproduce the kinetic energy reasonably accurately over all time steps considered and all four integrators reproduce the potential energy to within 1\%. While R2 and R4 are more accurate than the Euler and Verlet algorithms, they also require more CPU time per time step; to perform a single time step using R4 requires that the viscosity matrix $K_{pq}$ and elasticity vector $E_p$ be recalculated 4 times. In practise, the Euler integrator often offers the best accuracy to computational expense ratio, since an error of $~1\%$ is tolerable for most applications.

\begin{figure}[!]
\caption{{Convergence of the kinetic and potential energy averages as a function of the time step of a 54 element cylindrical mesh, where the unit time step is $\Delta t_0$=0.0001. This graph displays that as the time step decreases the error in the average energy in each quadratic degree of freedom tends to zero.}}
\begin{center}
\includegraphics[width=1.0\textwidth]{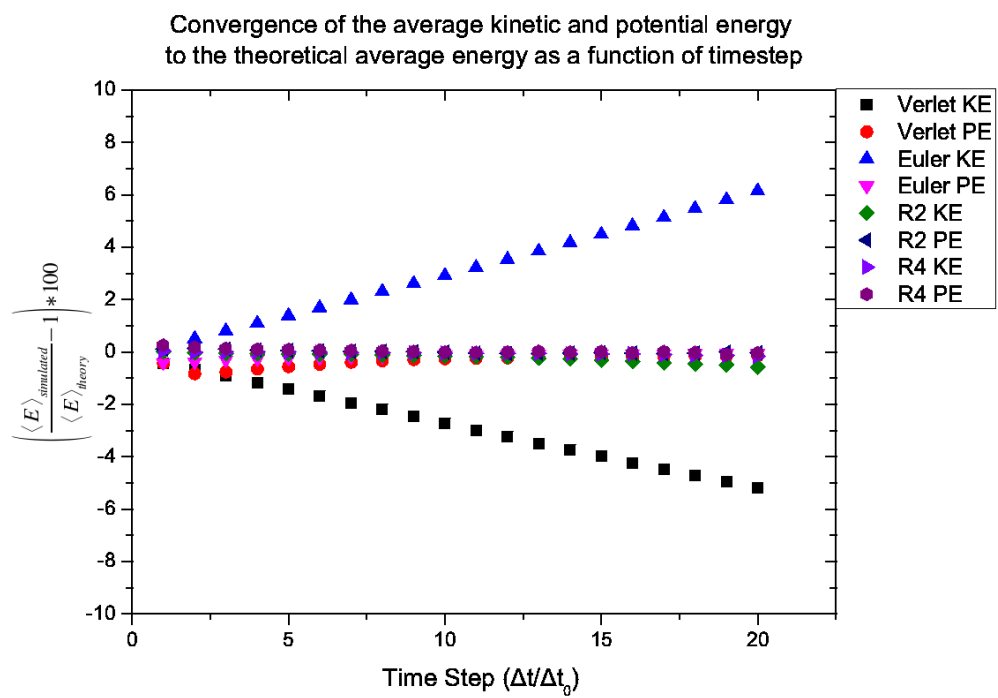}
\end{center}
\end{figure}

\subsubsection{Euler Beam Theory} The energy convergence tests in Section 3.1 show that the fluctuation dissipation relation derived in Section 2 is obeyed and that the correct theoretical averages for the kinetic and potential energies are obtained. These tests do not on their own show that the set of deformations predicted by the model are statistically correct. In order to test the conformational dynamics predicted by the FFEA model we consider the flexing of a thin rod due to thermal fluctuations, and compare the vibrational modes this system sustains from those derived from Euler Beam Theory [24]. For a classical beam undergoing pure bending, the equilibrium deflection $h$ due to an external torque $\tau$ is given by:

\begin{equation}
\frac{d^2 h(x)}{dx^2} = \frac{\tau(x)}{EI},
\end{equation}

\noindent where $E$ is the Young's Modulus and $I$ is the second moment of inertia of the area. The product $EI$ is the flexural rigidity. Equation (45) holds for thin beams when the deflection $h(x)$ is small relative to the length. For a uniform torque $\tau(x)$ the solution to Equation (45) with boundary conditions $h(0)=0$ and $h(L)=0$ (where $L$ is the beam length):

\begin{equation}
h(x)=\frac{\tau x(L-x)}{2EI}.
\end{equation}

Since Equation (46) provides the solution of Equation (45) for beam undergoing bending due to an external torque, we can obtain the flexural rigidity $EI$ by applying an external torque to a beam with the thermal noise turned off in the model. The amount of work required to bend a beam according to the function $h(x)$ is given by:

\begin{equation}
W= \frac{EI}{2} \int^L_0 \left ( \frac{\partial h(x)}{\partial x} \right )^2 dx.
\end{equation}
If $\tau=0$ at both ends of the beam so that $\frac{\partial^2 h}{\partial x^2}=0$, $h(x)$ can be expressed as a Fourier sine series:
\begin{equation}
h(x)= \sum_p h_p \sin \left ( \frac{p \pi x}{L} \right ).
\end{equation}
Substituting Equation (48) into (47) gives the amount of work done in each of the mutually orthogonal Fourier modes that correspond to a degree of freedom of the system, so that:

\begin{eqnarray}
W= \sum_p W_p,
\end{eqnarray}
where:
\begin{eqnarray}
W_p&=& h^2_p \left (\frac{EI}{4} \right )\left (\frac{p \pi}{L} \right )^4 L.
\end{eqnarray}
Therefore, from equipartition of energy it follows that if the beam is subject to thermal fluctuations then:

\begin{eqnarray}
\langle W_p \rangle &=& \frac{k_B T}{2}.
\end{eqnarray}

\subsubsection{Numerical Calculations for Beam Bending}We tested a total of eight finite element meshes; three have a hexagonal cross-section, four octagonal and one square (See Figure 2). {The hexagonal and octagonal beams have a maximum radius of 10nm, the square beam has sides of length 10nm and all beams  have a total length L of 160nm. In all simulations the viscosities were set to 3 times that of water, the elastic moduli $G$ and $B$ were set to $10MPa$, giving a Young's modulus of $20MPa$ and the density used was that of water. Thus, these simulations reproduce the thermal fluctuations of a hypothetical "nanogel" beam.}  The numerical tests are divided into two sections. First, we determine the flexural rigidity $EI$ of the beams. {This tests the influence of the mesh resolution,  and in addition investigates the effect of different finite element meshes. Secondly, we obtain the average energies in the first and second Fourier modes
to confirm that the deformations of the beams obey the correct statistics.}

\begin{figure}[!]
\caption{The eight beam meshes used to test configurational fluctuations in FFEA. Only the surface meshes of Hex2 and Hex3 are the same, internally the element structure is different. Similarly, for Oct2 and Oct3 the internal nodes are placed slightly differently to ensure the the results obtained are independent of the arrangement of finite elements. Oct4 is the beam mesh used to perform the fine grained calculations in Section 3.2.3 and the square beam mesh is used in the second order element scheme described in Section 3.2.4. For the square beam mesh, all the linear elements in the system within the second order element structure are shown.}
\begin{center}
\includegraphics[width=1.0\textwidth]{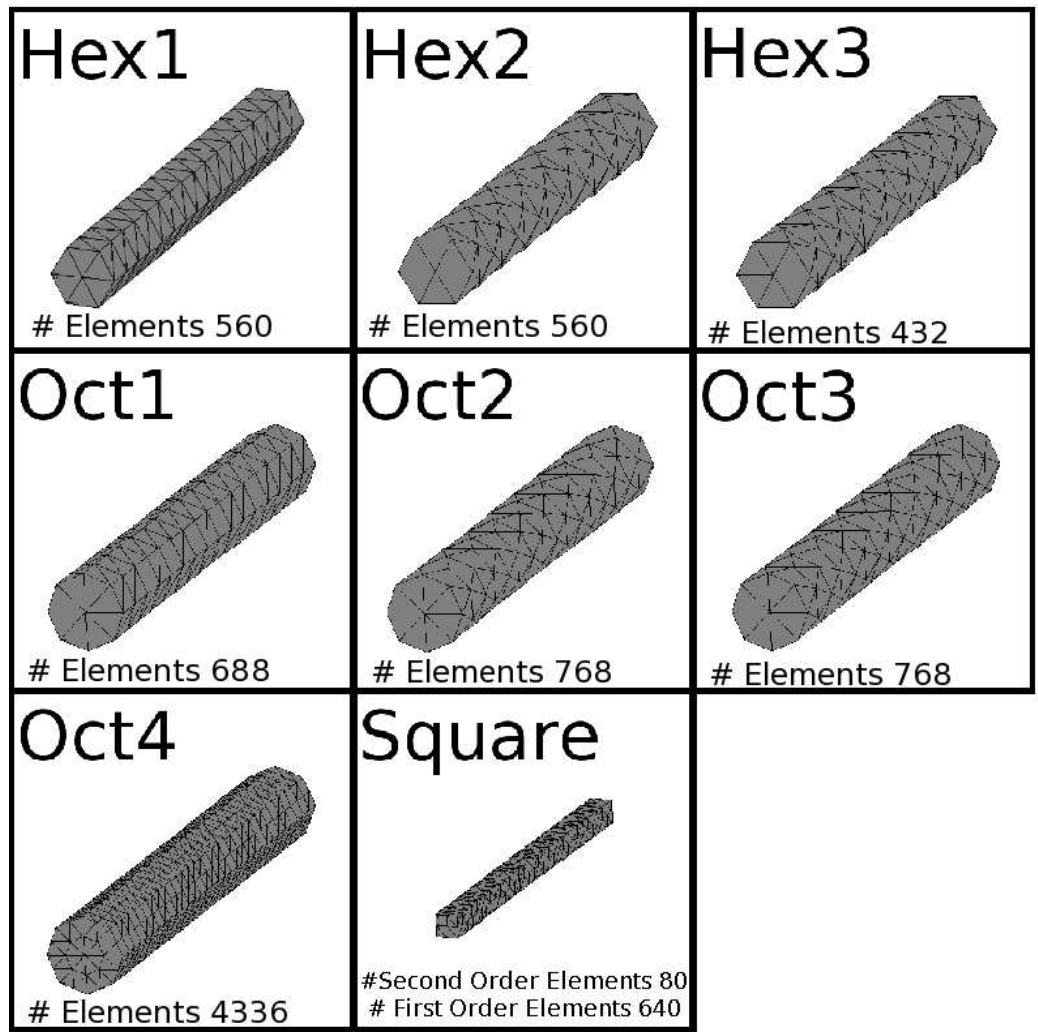}
\end{center}
\end{figure}

\pagebreak

To determine the flexural rigidity an external torque $\tau$ was applied to the end of each beam and the thermal noise term was switched off. {Prior to finite element discretisation, the governing continuum equation for this system is:}

\begin{equation}
{\rho \left ( \frac{\partial u_i}{\partial t} +u_{j}\frac{\partial u_i}{\partial x_j} \right ) = \frac{\partial \sigma^{v}_{ij}}{\partial x_j} + \frac{\partial \sigma^{e}_{ij}}{\partial x_j}+ \tau_i}.
\end{equation}

\noindent {To impose a torsional stress on each beam, a torque $\tau$ of magnitude $\tau=5x \cdot 10^{-20}Nm$ (where $x$ is the linear distance from the central axis of the beam) was applied to the end faces only. Stress free boundary conditions were used elsewhere.} The results of these eight calculations are presented in Table 1.

\begin{table}[ht]
\caption{Flexural Rigidity Results for Different Meshes}   
\centering                          
\begin{tabular}{|c|c|c|}            
\hline\hline                        
Beam & $\frac{(EI_x)_{Simulated}}{(EI_x)_{Theory}}$ & $\frac{(EI_y)_{Simulated}}{(EI_y)_{Theory}}$  \\ [0.5ex]   
\hline                              
Hex1 & 1.70 & 1.70  \\               
Hex2 & 1.60 & 1.60  \\
Hex3 & 1.82 & 1.75  \\
Oct1 & 1.61 & 1.50  \\
Oct2 & 1.48 & 1.48  \\
Oct3 & 1.48 & 1.48  \\
Oct4 & 1.30 & 1.26  \\
Square & 1.00 & 1.00  \\ [1ex]         
\hline                              
\end{tabular}
\label{table:bending}          
\end{table}

For linear finite elements the flexural rigidity of the long thin beams is larger than predicted theoretically. This is a consequence of there being only a small number of linear elements across each cross-section, which artificially stiffens the rods. We have devised two solutions to correct this over estimation of the flexural rigidity for linear finite elements. {One solution (discussed in Section 3.1.4) utilises a refined linear mesh with more finite elements spanning the diameter of the beam thus improving the interpolation of displacements}. The second  (discussed in Section {3.1.5}) uses second order elements to describe the displacements and elastic stresses.


{Now that the flexural rigidity of each beam has been obtained (see Table 1), the thermal noise is reintroduced so that the Fourier modes can be extracted. The temperature of the system was set to be 300K.} To maintain small deformations, $\frac{k_B T L}{EI}$ is set to be approximately $10^{-3}$. {With stress free boundary conditions everywhere, the governing equation for this simulation prior to finite element discretisation is now given by:}

\begin{equation}
{\rho \left ( \frac{\partial u_i}{\partial t} +u_{j}\frac{\partial u_i}{\partial x_j} \right ) = \frac{\partial \sigma^{v}_{ij}}{\partial x_j} + \frac{\partial \sigma^{e}_{ij}}{\partial x_j}+\frac{\partial \sigma^{t}_{ij}}{\partial x_j}}.
\end{equation}

{For the Hex1-3 and Oct1-3 beams, a total of 21 independent repeat simulations were performed. This ensured sufficient sampling of the first and second Fourier modes in the x and y directions perpendicular to the beam axis. There are two important timescales in this system; the oscillatory timescale for the first harmonic and the decay time of these oscillations. The longest oscillatory time scales for the Hex1-3 and Oct1-3 beams are of order $10ns$; the longest decay time scale for these oscillations was measured to be around $10ns$. Since the total simulation time was $1.5 \mu s$, both of these important time scales were adequately sampled. Each Fourier amplitude was then averaged, the  variance of the distribution obtained; substitution into Equation (50) then provides the average energy of that particular Fourier mode (see Table 2).}


\begin{table}[ht]
\caption{Average energies in the first and second Fourier modes normalised by $ \frac{k_B T}{2}$ (so that the correct theoretical answer is 1). FM1X and FM1Y denote the average energy in the first and second Fourier modes in the X direction directions respectively, while FM2X and FM2Y refer to the average energy in the second Fourier modes. Where the error is given by the standard deviation of the resultant distribution of energies from each of the 21 different simulations.}
\centering                          
\begin{tabular}{|c|c|c|c|c|}            
\hline\hline                        
Beam & FM1X $\left (\frac{2 \langle W_p \rangle}{k_B T} \right )$ & FM1Y $\left (\frac{2 \langle W_p \rangle}{k_B T} \right )$ & FM2X $\left (\frac{2 \langle W_p \rangle}{k_B T} \right )$ & FM2Y $\left (\frac{2 \langle W_p \rangle}{k_B T} \right )$ \\ [0.5ex]   
\hline                              
Hex1 & 1.014 $\pm$ 0.068 & 1.058 $\pm$ 0.038 & 1.000 $\pm$ 0.034 & 1.024 $\pm$ 0.032\\               
Hex2 & 0.940 $\pm$ 0.064 & 0.982 $\pm$ 0.056 & 0.960 $\pm$ 0.034 & 0.966 $\pm$ 0.030\\
Hex3 & 0.976 $\pm$ 0.044 & 1.004 $\pm$ 0.062 & 0.910 $\pm$ 0.028 & 0.924 $\pm$ 0.024\\
Oct1 & 0.984 $\pm$ 0.048 & 0.928 $\pm$ 0.052 & 1.022 $\pm$ 0.030 & 1.052 $\pm$ 0.028\\
Oct2 & 1.026 $\pm$ 0.082 & 0.982 $\pm$ 0.062 & 1.046 $\pm$ 0.036 & 1.064 $\pm$ 0.042\\
Oct3 & 1.010 $\pm$ 0.076 & 0.906 $\pm$ 0.070 & 0.974 $\pm$ 0.044 & 0.980 $\pm$ 0.034\\ [1ex]         
\hline                              
\end{tabular}
\label{table:EnergyFM}          
\end{table}

The results for the different meshes using the flexural rigidities from Table 1 all show good agreement with the theoretical prediction for the average energy in the first and second Fourier bending modes. The results agree with the theoretical average energy predicted by the equipartition theorem within the calculated sampling error.

Figure 3 shows {four} representative conformations of the beams sampled from the FFEA simulations. These were obtained by plotting the centre of mass of different sub-sections of the beam along its length relative to the beam ends to represent the instantaneous configuration (consequently the ends of the beams always have a total displacement of zero). {Furthermore, the deflections of each centre of mass of the beam follow a Gaussian distribution as expected for a beam subject to thermal noise.}

\begin{figure}[!]
\caption{{Four different} conformations adopted by the Hex1 beam due to thermal noise. The boundary condition of no external torque $\frac{\partial^2 h}{\partial x^2}=0$ enables the deformations $h(x)$ to be measured relative to the positions of stationary beam ends.}
\begin{center}
\includegraphics[width=1.0\textwidth]{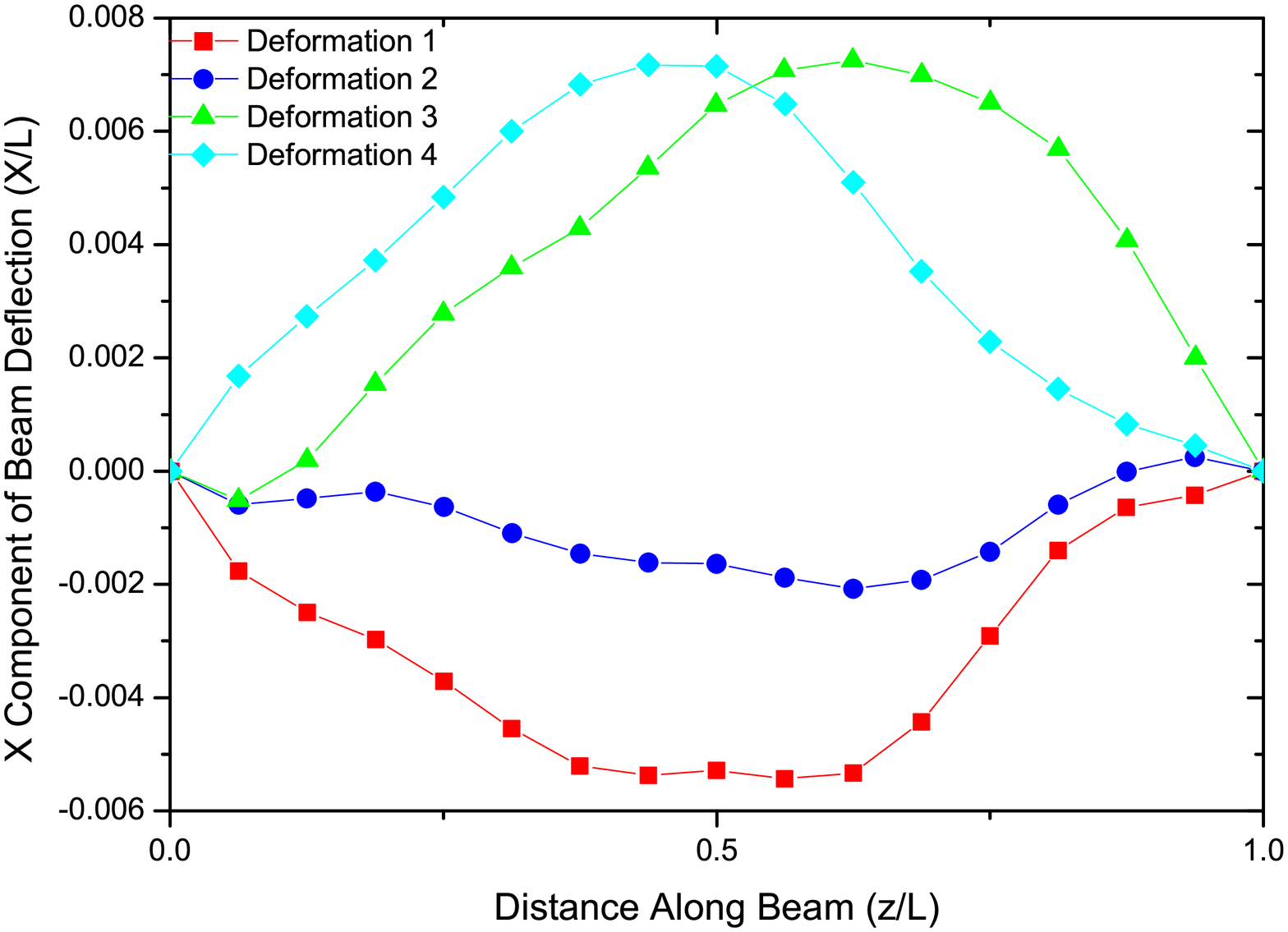}
\includegraphics[width=1.0\textwidth]{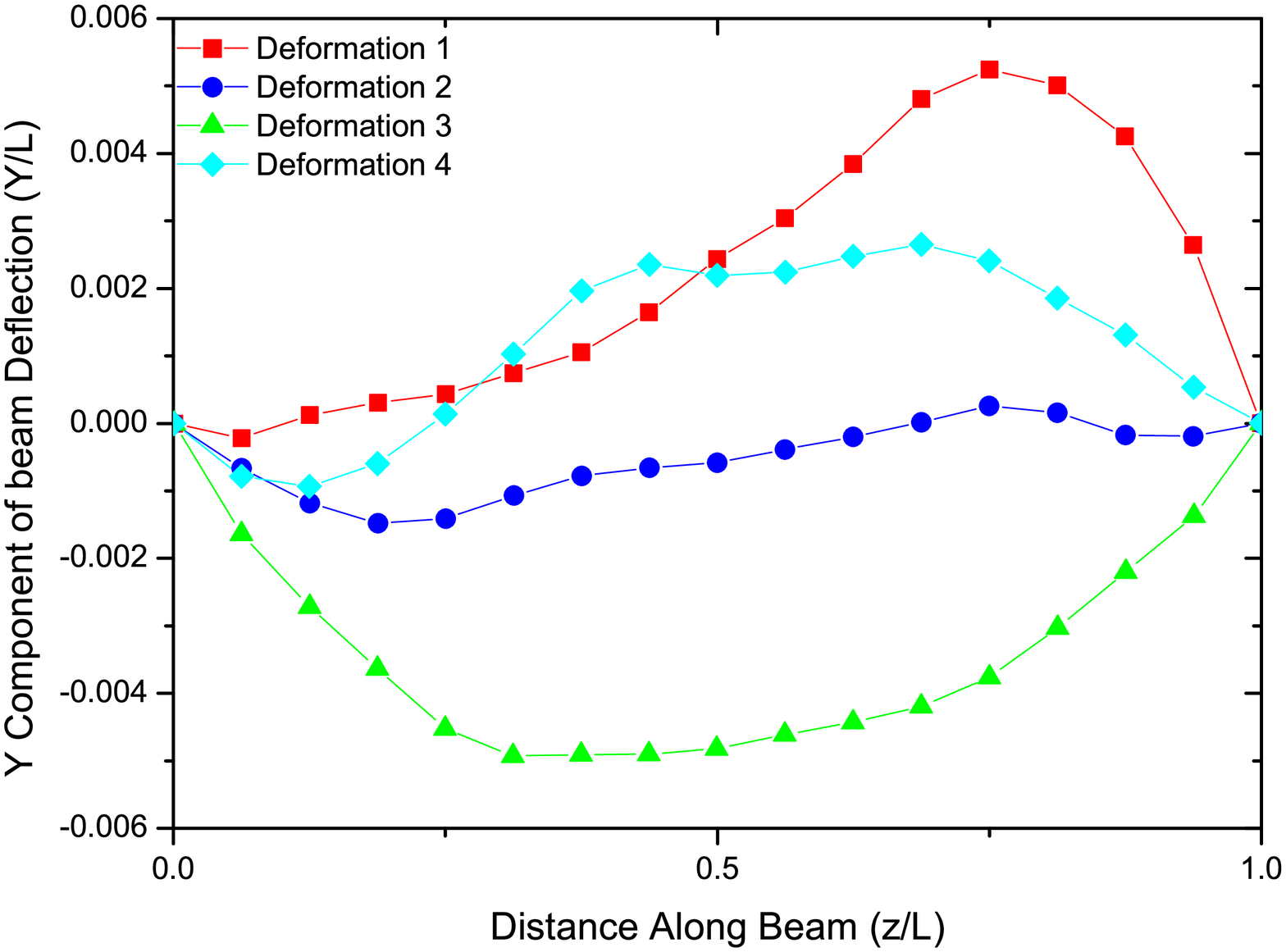}
\end{center}
\end{figure}

\pagebreak

From the calculation of the flexural rigidities from the finite element meshes of the {eight beams (Table 1)}, and the calculation of the average energies in the first two Fourier modes, {we conclude that the average energies obtained from FFEA are in agreement with theoretical predictions}. However, the linear approximation for the elements can lead to an over-estimation of the flexural rigidity when the mesh contains too few elements. To demonstrate that this can be corrected, we have firstly performed simulations which retain linear elements but which employ a finer mesh, and secondly we have derived a second order element scheme for the elasticity.

\subsubsection{Fine Grained Mesh} To capture the bending of the beams more accurately we calculated the flexural rigidity using an octagonal mesh with four elements across the diameter of the beam (Oct 4) compared to the two used in Oct1, 2 and 3. The same {viscosities, elastic moduli} and density were used as previously. As shown in Table 2, improving the mesh resolution halves the error in the flexural rigidity measured. However, this solution is more numerically costly as there are approximately 8 times as many elements to be considered in this finer grained mesh.

\subsubsection{Second Order Element Solution} As expected, increasing the mesh resolution does indeed improve the flexural rigidity. An alternative method is to modify the FE algorithm to give a more accurate treatment of the elasticity by including quadratic terms in the interpolation of displacements. This requires a solution of Equation (9) in which the elastic terms and the mass matrix are solved using second order elements [25]. The elastic stress is calculated using 10 node isoparametric tetrahedral elements from which Equation (12) can be solved using second order shape functions. In general, this integral cannot be performed analytically since in the second order regime the local compression within an element is not isotropic. Thus, the integrals need to be performed numerically using Gaussian quadrature. As discussed in Section 2.3.2, it is much more difficult to include viscous and thermal noise terms for a second order finite element mesh. 
We therefore retain a linear solution for the thermal and viscosity terms by subdividing each quadratic element into 8 linear elements, and using these sub-elements to calculate the viscous and thermal noise terms. The viscosity matrix and thermal force vector are calculated by subdividing each isoparametric tetrahedron into linear elements and then performing the integrals in Equations (11) and (13) for each of the linear sub-elements.

To test the quadratic element solution, we repeated the beam bending calculations and obtained the flexural rigidity for a simple square cross-section beam with side length 1. As shown in Table 1, the second order elements give the expected flexural rigidity for a square cross-section to within the accuracy of our measurements. The use of second order elements provides a more accurate solution than the increasing the number of linear elements, even when the linear mesh resolution is improved by a factor of 8. Since the main increase in the computational expense for the quadratic elements arises from calculating the thermal and viscosity terms, which involve the contributions from the eight linear sub-elements that make up each quadratic element, the second order solution gives better efficiency in the trade-off between accuracy and computational expense.

\section{Application to Protein Modelling} Finally, we demonstrate the use of FFEA to model the conformational flexibility of a globular protein using the first order element approximation to improve computational efficiency. As a representative system, we have used the long fatty acid chain Co-A ligase enzyme from the organism \emph{Fusobacterium nucleatum}. To date, it has not been possible to obtain atomically detailed structural information for this protein. However, the overall 3-dimensional shape of the biomolecule has been determined using Small Angle X-ray Scattering (SAXS) [26]. Figure {4(a)} shows the atomistic structure of the homologous protein Archaeoglobus Fulgidus (PDB ID: 3G7S [27]), with the SAXS structural envelope of the Co-A ligase superimposed. The experimentally determined structural envelope was converted into a finite element mesh using TETGEN [28], which was then further refined using NETGEN [29].
The resulting mesh is compared with the original SAXS structural envelope in Figure {4(b)}.

\begin{figure}[!]
\caption{Comparison of the SAXS structural envelope and atomistic structure of the homologue X with the equivalent finite element mesh viewed in NETGEN}
\begin{center}
\includegraphics[width=1.0\textwidth]{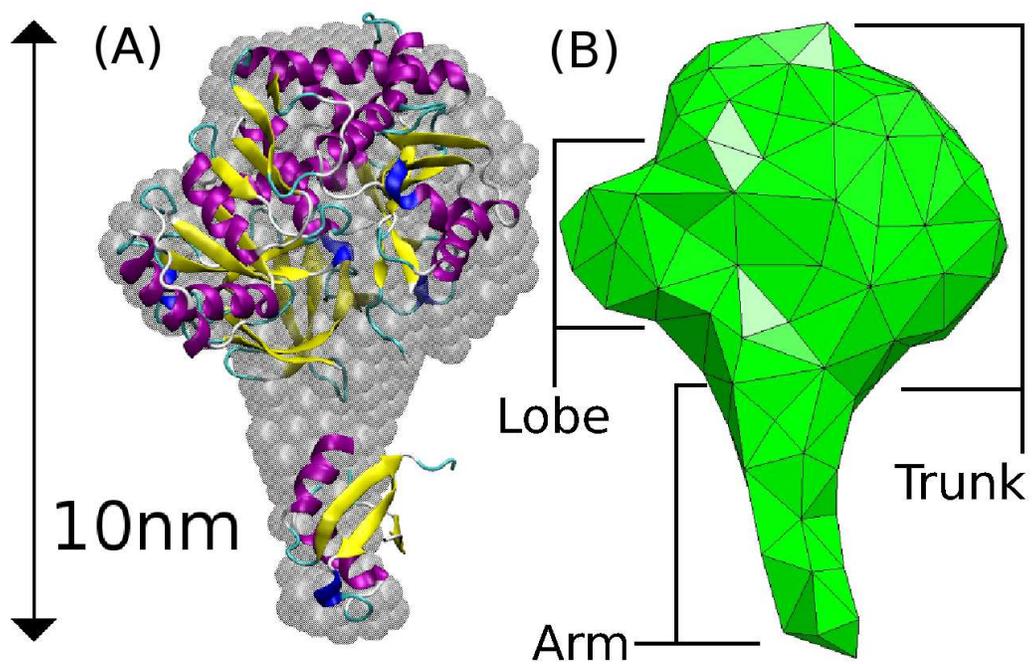}
\end{center}
\end{figure}

The material properties of the biomacromolecule were assigned based on the existing literature values quoted for proteins, where available. The density of a globular protein [30, 31] is around $1500 kg m^{-3}$. Currently very little is known about the internal viscosity of proteins. Therefore, we have assumed a value that corresponds to the shear viscosity of water, namely $10^{-3} Pa \ s$. The same value was used for the bulk viscosity  [32]. The temperature was set to $300K$. The Young's modulus of a number of proteins has been measured using Atomic Force Microscopy, and values of between $40 MPa$ and $5 GPa$ [33, 34, 35, 36] have been reported. Given that the Poisson's ratio for a biomolecule has been estimated to be around 0.4 [36], it is possible to assign the bulk and shear moduli corresponding to a particular choice of Young's modulus. 
For these calculations, we tested the model using three values of the Young's modulus corresponding to low ($450MPa$), medium ($560MPa$) and high ($800MPa$) biomolecular flexibility. The bulk and shear moduli can then be calculated from the following, where $E$ is the Young's Modulus and $\nu$ the Poisson ratio:

\begin{eqnarray}
G &=& \frac{E}{2(1+\nu)} \\ K &=&\frac{E}{3(1-2\nu)}
\end{eqnarray}

{Prior to finite element discretisation, the governing equation for the protein model is given below in Equation (56):}

\begin{equation}
{\rho \left ( \frac{\partial u_i}{\partial t} +u_{j}\frac{\partial u_i}{\partial x_j} \right ) = \frac{\partial \sigma^{v}_{ij}}{\partial x_j} + \frac{\partial \sigma^{e}_{ij}}{\partial x_j}+ \frac{\partial \sigma^{t}_{ij}}{\partial x_j}.}
\end{equation}

{Using the governing Equation (56), with stress free boundary conditions and the mesh shown in Figure {4(b)}, FFEA analysis was performed for 500$ns$ for the three different choices of material parameters.} Each calculation had a run time of around 2 weeks on a single CPU. The simulation trajectories were visualised using paraFEM [37]. On visualising the trajectories, it was apparent that the molecule changes its orientation relative to the starting structure, whilst conserving angular momentum, as the fluctuations in the shape of the biomolecule cause the inertia tensor to change [38]. Therefore, the trajectories were post-processed to reorientate the molecule prior to analysis. The biomolecular flexibility was quantified by calculating the Root Mean Squared Deviation (RMSD) of the co-ordinates of the mesh nodes from their initial values during the simulations, as shown in Figure 5. 
As expected, increasing the Young's modulus from $450 MPa$ (red line) to $800 MPa$ (blue line) results in a smaller RMSD from the initial structure, indicating a less flexible protein. Given that a series of $10ns$ atomistic MD simulations of small proteins selected from the protein data bank obtained RMSD values of between 1 and 6{\AA}, the values that we obtain (2\AA) are reasonable [39-42].

\begin{figure}[!]
\caption{RMSD obtained for three different sets of elastic parameters (with rotations removed prior to analysis).}
\begin{center}
\includegraphics[width=1.0\textwidth]{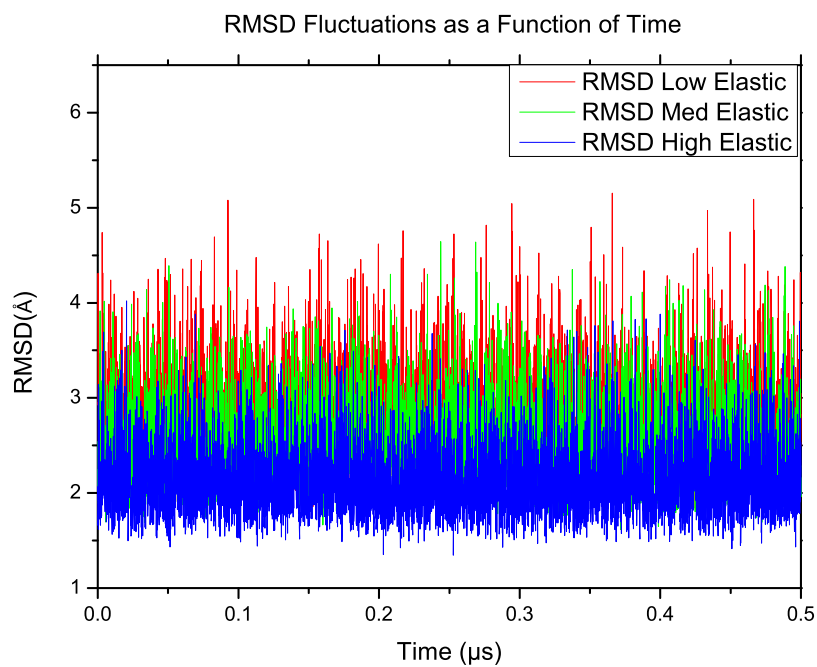}
\end{center}
\end{figure}

\pagebreak

In an analogous manner to conventional particle-based molecular dynamics, but now at the continuum level, FFEA provides a series of conformers of the protein as it undergoes thermal fluctuations. Figure 6 shows 9 representative conformations of the protein extracted from the FFEA simulation trajectories performed with the lowest Young's Modulus ($450 MPa$).

\begin{figure}[!]
\begin{center}
\includegraphics[width=1.0\textwidth]{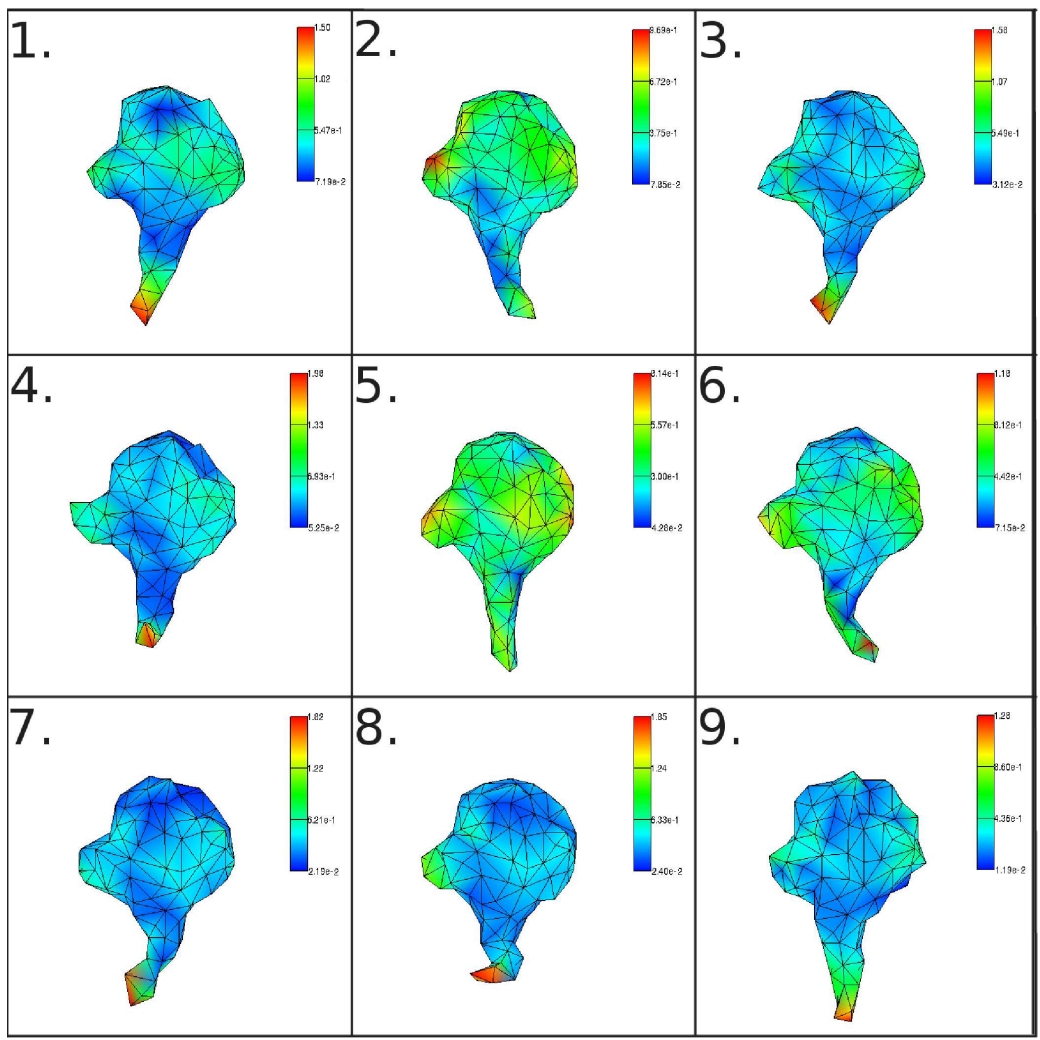}
\end{center}
\caption{Nine representative conformers Co-A ligase sampled from FFEA simulations with $E=450 MPa$: 1. Arm swings left. 2. Small thermal disruptions to the entire trunk, including lobe.  3. Large arm swing to the left and lobe disruption. 4. Arm sticks out of the page and disruption to the leftward lobe. 5. Entire protein elongated by the thermal noise. 6. Arm swings to the right. 7. Arm swings to the left and major disruption to the shape of the arm. 8. Arm swings out of the page. 9. Elongation with change in shape of the entire trunk.}
\end{figure}

\pagebreak

Each of the 9 snapshots are coloured by their overall deviation from the equilibrium configuration with the scale bar showing the displacement in nanometers. Figure 6 shows that the "trunk" of the protein is relatively immobile, and undergoes minor structural disruptions while retaining its overall shape. However, the lobe located to the left of the biomolecule moves more significantly than other regions of the trunk as it is considerably thinner than the main body of the protein. The most striking deformations occur in the "arm" at the base of the enzyme, which is highly flexible and swings back and forth around the bottom of the molecule during the course of the simulation. This indicates that the intermediate region between the arm and the trunk acts as a flexible hinge region in the biomolecule. It is interesting to note from Figure {4(a)} that is it precisely in the region that FFEA predicts should be of greatest flexiblity that the homologous protein Archaeoglobus Fulgidus has missing electron density, indicating that this region was too mobile for its structure to be determined crystallographically.

\section{Conclusions}We have developed an extension of Finite Element Analysis that includes thermal fluctuations, and which thereby extends the applicability of this technique from macroscopic materials to soft nano-scale objects, such as biomacromolecules. This reduction in scale into the nanometer regime is achieved using a local fluctuating thermal stress to apply thermal agitation  to a {continuum material consistent with a Kelvin-Voigt model}. The localised nature of the thermal noise avoids the need to invert a global resistance matrix. The algebraic derivation of the noise term has been validated by demonstrating numerically that the algorithm applied to a simple linear beam provides potential and kinetic energies in agreement with the fluctuation-dissipation theorem. Similarly, we have demonstrated that the energy in the first two Fourier modes of a long thin beam is in agreement with the equipartition theorem.

Given an overall shape of a macromolecule, and appropriate estimates for its material properties (such the density, bulk/shear moduli and bulk/shear viscosities), fluctuating finite element analysis can provide a continuum dynamics trajectory describing the changing shape of the macromolecule as it undergoes thermal fluctuations. The algorithm has the advantage over many of the existing biomolecular modelling techniques, such as conventional atomistic simulations, in that it does not require a detailed atomic structure as input to the calculation. Consequently, we have been able to apply FFEA to the Co-A Ligase enzyme from the organism \emph{Fusobacterium nucleatum}, for which it has only been possible to obtain lower resolution structural information to date. Assuming physically realistic values for the material parameters of the protein gives RMSD values from equilibrium that are consistent with estimates obtained for proteins from atomistic simulations [41]. 
However, we have shown that changing the input material parameters, specifically the Young's modulus, changes the magnitude of the thermal disruptions; stiffer proteins with a larger Young's modulus undergo smaller thermal fluctuations. It is extremely difficult to obtain atomically detailed experimental information on biomolecular flexibility. However, developments in lower resolution techniques, such as SAXS [43], cryo-electron microscopy [44], fluorescent resonance energy transfer labelling [45], ion-mobility mass-spectroscopy [46] and atomic force microscopy [47] are all starting to provide dynamic information at the mesoscopic level. FFEA uniquely offers the possibility of using computational methods to quantitatively assign mesoscale material parameters to individual biomolecules from such experimental data by systematically varying the input parameters until they match the experimental results. Such an analysis would allow for a quantitative comparison of the flexibilities of biomacromolecules, which explicitly takes
the shape of the molecular envelope into account. For molecules where the structure is known in atomistic detail, this will improve our understanding of the relationship between atomic structure and global flexibility. For biomolecules which must be inherently dynamic in order to perform their function, such as molecular motors, we will gain new insight into the physics underlying their mechanism of action.

By providing trajectories at the mesoscopic continuum level, FFEA has the potential to provide information complementary to that of detailed atomistic simulation, but at the next level up in terms of system size and simulation timescale. Although our focus is on biomolecular dynamics, FFEA could equally well be applied to the meso-modelling of microgels [48], block copolymers [49] or soft colloids [50]. The algorithm has been designed to be sufficiently flexible that it will accommodate successive methodological developments to both improve its accuracy for protein modelling and to broaden the range of biological questions that can be addressed. We anticipate that introducing inhomogeneous material parameters within a protein of known secondary structure will enable us to provide a more accurate representation of biomolecular dynamics. 
This is straightforward, given that there is no requirement for the finite element mesh to be homogeneous. Moreover, techniques for calculating continuum material properties from atomistic structures of biomolecules have already been reported [51]. We are currently modifying the stress tensor (Equation (2)) and utilising the boundary element method to include short range repulsion between macromolecular surfaces, attractive dispersion forces, electrostatics and {exterior hydrodynamics} so that FFEA can be applied to protein-protein complexes, protein-surface interactions and crowded macromolecular environments.

\section{Acknowledgements} Robin Oliver was funded through an EPSRC DTA studentship. We thank Thomas Grant, Edward Snell and Joesph Luft for providing the experimental SAXS envelope for Co-A Ligase [26].





\bibliographystyle{model3a-num-names}


\end{document}